# LINEAR AND NON-LINEAR BARRIER COVERAGE IN DETERMINISTIC AND UNCERTAIN ENVIRONMENT IN WSNs: A NEW CLASSIFICATION

ADDA BOUALEM[1], DJAHIDA TAIBI[1] AND AROUA AMMAR[1]

[1]Department of Science Computing, Ibn Khaldoun University, Tiaret, Algeria
adda.boualem@univ-tiaret.dz, tiabidjahida05@gmail.com, ammararoua20@gmail.com

## ABSTRACT

*This paper Various studies cited in the literature deal with the classic problem of obstacle coverage, where the deployment environment, sensor nodes, and base stations have characteristics that are considered perfect but suffer from various flaws in the real world. This paper presents other barrier coverage types ranked in a new classification based on linear and nonlinear barrier coverage according to deterministic and insecure environments, and enumerates some of the different current and future challenges of these coverage types and connectivity in WSNs.*

## 1. INTRODUCTION

The Wireless Sensor Network (WSN) technology was born with the recent advances in wireless communication and electronics that have enabled the development of inexpensive, low-power, and versatile sensors that are small enough to operate over short distances communicate within. The real mode is merged by the different types of uncertainties, such as imprecision in measurement components, atmospheric phenomena, intrusion, animals, and natural phenomena such as volcano, rivers, industrial phenomena and others [1]. The unreliability of communication radios and reception radios, etc., affects the quality of service and decisions regarding real-world information [2].

This survey addresses the problems of Linear-based barrier coverage models and Non-Linear-based barrier coverage models. This purpose is to expand the wireless sensor network as much as possible to overcome the previous uncertainties and guarantee the quality of service. Thus, the aim is to ensure barrier coverage with a minimal number of connected node subsets, and dominant nodes and minimal cost, regardless of the type of deployment (random or deterministic).

This paper is structured as follow; Section 2 reviews the problems that influence the Wireless Sensor Networks, and show the relationship between these problems. Section 3 summarizes the different types of coverage proposed in the literature.

Section 4 cites some related works in barrier coverage. It is subdivided into two main parts. The first part presents some deterministic and uncertain-based coverage strategies and show the drawbacks to open up avenues for research. The second part focused on the new barrier coverage classification (the linear and non-linear types barrier coverage in deterministic and uncertain environment in WSN, and quotes some linear and non-linear- bases coverage strategies and show the drawbacks to open up avenues for research. Section 5 describes the current challenges and future challenges of linear and non-linear coverage. Finally, Section 6 presents conclusions and future work.

## 2. PROBLEMS THAT IMPACT WIRELESS SENSOR NETWORKS

Implementing sensor networks poses many challenges for researchers to meet stringent constraints imposed by certain properties, including:
- Energy sources are very limited,
- An unattended and harsh deployment environment (the hostility of the environment),
- limited and unsecured radio links (unreliable communication),
- Changing network topologies (changing topologies of deployed sensor networks),
- Uncertain characteristics of either the deployment environment, or the components of the sensor nodes used,
- Heterogeneity of nodes and multi-hop communications, etc.

The most problems that affect the WSn are; Energy, Node Cost, Limited bandwidth, Deployment, Security, etc.

The Fig 1 cites the majority of known problems in the literature.

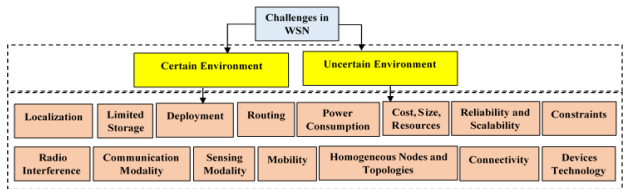

Figure 1. WSN Problems [3]

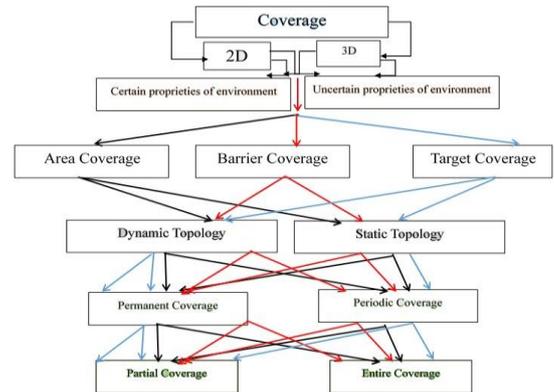

Figure 2. WSN Coverage Types [3]

## 3. COVERAGE TYPES

There is a range of types proposed in the literature, as illustrated in Figure. 2. The main types are:
- **Target coverage:** Target Coverage, also known as Point Coverage, is used to monitor specified targets in an AoI.
  Fig. 3.(1) illustrates a target coverage scenario where six sensor nodes are deployed to monitor the five targets in the Area of interest (AoI). Where, target $t_5$ is covered by only one sensor node. The other targets; $t_1$, $t_2$, $t_3$ and $t_4$ are simultaneously covered by two sensor nodes. Target coverage reduces energy consumption because it monitors
  Only specifies the position of the target within the AoI.
- **Area Coverage:** The area coverage, is to monitor the AoI by a minimal set of deployed sensor nodes in the area of interest (Figure.3.(3)).
- **Barrier Coverage:** The Barrier Coverage is to deploy a limited number of sensing nodes in area borders, for monitoring, and building a Barrier against intrusion (Figure 3.(2)).

A comparison between the different strategies used for k-barrier Coverage according to different classifications in Table 1 and Table 2.

## 4. NEW CLASSIFICATION OF BARRIER COVERAGE

The barrier coverage of a geographical area by a network of sensors focuses on the answers to the questions:
- How to cover a geographical area against all intrusion types with a minimal set of sensor nodes?
- How can coverage and connectivity be guaranteed for a maximum period for applicable objectives?
- How can coverage be guaranteed for a maximum period of time in uncertain environment?
- In which deterministic and uncertain positions should sensor nodes be placed or deployed to ensure maximum coverage of the area of interest?

In practice, barrier coverage is usually embodied in the answer to two basic questions [16] :
- How to evaluate barrier coverage performances rate, when sensor nodes are deployed in form of monitoring barriers?
- How to improve coverage performance when the active barriers is minimal?.

Many researchers are currently interested in developing solutions that address various needs [17], and have proposed divers classification such the classification illustrated in Figure. 4.

The Barrier Coverage aims to detect and to minimize the intrusion through the network (WSNs). Boulis [18] pointed out that the monitoring of WSN is based on the management of three factors: the energy level of each sensor node, the coverage area and the connection characteristics between nodes. [11]

Intrusion detection is one of the most important issues in WSNs, whose goal is to detect any intruder trying to enter the network. In fact, many security applications need to detect intruders, such as border protection, critical infrastructure protection and hazardous substance monitoring. In WSNs, area coverage and barrier coverage [17] are proposed to achieve the objective of intruder detection. In this case, we proposed a new classification of barrier coverage types; (a) Linear barrier coverage, and (b) non- linear barrier coverage. In this case. Each type characterized by strong or weak barrier, in static or dynamic topology, depending on the need for permanent or periodic coverage of the application.

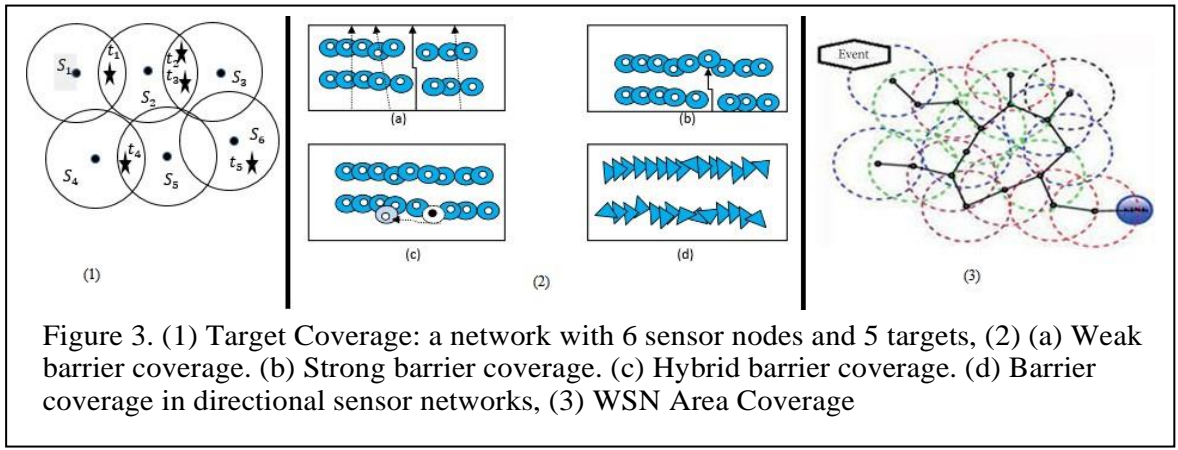

Figure 3. (1) Target Coverage: a network with 6 sensor nodes and 5 targets, (2) (a) Weak barrier coverage. (b) Strong barrier coverage. (c) Hybrid barrier coverage. (d) Barrier coverage in directional sensor networks, (3) WSN Area Coverage

| Ref/ No, Pub Year | Paper Highlight | Deployment Type | Sensor type | Connectivity | K-Barrier Type | Space |
|---|---|---|---|---|---|---|
| Hong Y, et al [4], 2022 | Monitor intruders in high resolution and maximize network longevity | Deterministic deployment | Cameras | Connected barriers | Strong Barrier | 3D |
| Zhanwei Yu [5], 2019 | Minimizes hybrid access point (H-AP) power supply by jointly optimizing power/information transfer timing and H-AP transfer performance while meeting node throughput requirements. | Random deployment | Static | Connected nodes | Weak Barrier | 2D |
| Ernest Bonnah et, al [6], 2020 | Coverage maximization scheme, by using the exposure profile from the sensor node to the sink node to calculate the minimum exposure path | Random deployment | Static | Connected nodes | Strong Barrier | 2D |
| Ernest B, et al [7], 2017 | Linear Programming (LP) techniques and on-linear optimization problems in WSN | Random deployment | Static | Connected nodes | Non-Linear Barriers | 2D |
| Kaiye G, et al [8], 2021 | Minimizing the cost of the entire system while meeting pre-specified requirements on the expected signal coverage and system reliability. | Random deployment | Mobile | Connected nodes | Linear Barrier | 2D |
| Hong Y, et al [4], 2022 | monitoring the intruder with high resolution and maximizing the network lifetime | Deterministic deployment | Cameras | Connected barriers | Strong Barrier | 3D |
| Zhan wei Yu [5],2019 | Minimize hybrid access point (H-AP) energy provision via jointly optimizing the time allocation of energy/ information transmissions and H-AP's transmission power while satisfying node throughput requirement. | Random deployment | Static | Connected nodes | Weak Barrier | 2D |

Table 1. Comparative study between the different k-Barriers Coverage works according the Connectivity and k-barrier Type.

| Ref/No, Pub Year | Paper Highlight | K-Barrier Type | Strategy/ Objectives |
|---|---|---|---|
| A. Boualem, et al [9], 2017 | Probabilistic interference and Truth-Table model for Strong K-Barrier Coverage | Strong Barrier | Sensing model |
| Xing-Gang Fan, et al [10], 2020 | An Efficient Scheme for Mobile Nodes to Join Barriers One by One to Form Strong Barrier Coverage | Strong barrier | Sensor mobility |
| Ernest Bonnah, et al [11], 2020 | The coverage maximization scheme uses the exposure profile of the sensor nodes from the sink node to calculate the minimum exposure path | Strong Barrier | Coverage intensity |
| Xiaoyang L, et al [12], 2018 | Genetic algorithm of route planning of WSN | Weak Barrier | Node localization method |
| A. Boualem, et al [9], 2017 | Probabilistic interference and Truth-Table model for Strong K-Barrier Coverage | Strong Barrier | Sensing model |

Table 2. Comparative study between the different k-Barriers Coverage works according k-barrier Type and strategy objective

## 5. Current and future challenges Barrier Coverage Problem

In many cases, inconsistent detection requirements must be considered when using WSNs, depending on the size or sensitivity of the surveillance area. For example, high detection accuracy is required for sensitive areas, and low detection accuracy is required for smaller areas. Atmospheric event influences shaping the physical environment  Location, communication strength and monitoring of sensor nodes in the network. This reality makes it necessary to consider the nature of uncertainty. Many proposed works in literature consists of introducing fuzziness in the process of scheduling sensor nodes in WSNs for several purposes. The following types of uncertainty are present in WSNs:

- Uncertainty in radio communication links. Communication performance increases with spatial distance. Mobility, power, performance, and connectivity are constraints that prevent network sensor nodes from communicating when deployed in a three-dimensional (3D) environment.
- Uncertainty in the testing process. Environmental disturbances, angles, nonlinear distances, noise, sensor types, and other factors will bring uncertainty to the detection process of sensor networks.
- Detection uncertainty in data collection. When sensors are deployed in harsh environments, various factors can affect the quality of the collected or captured data, such as degradation (wind, terrain, animals, etc.).

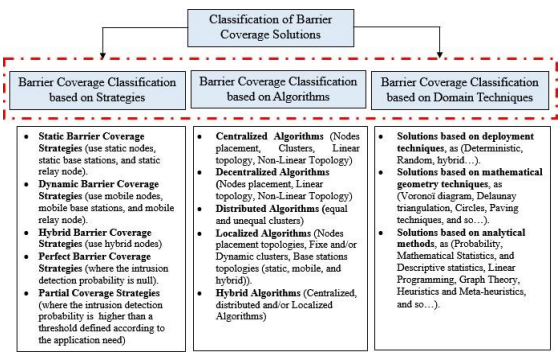

Figure 4. Our WSN Barrier Coverage

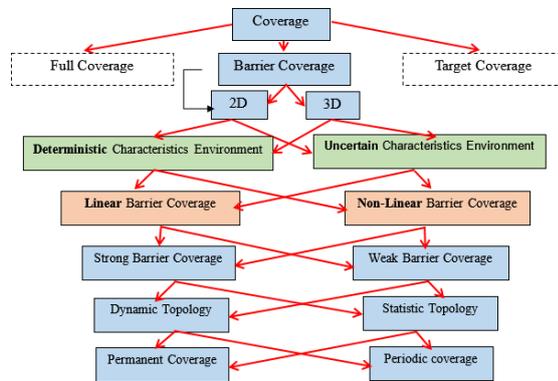

Figure 5. The New WSN Barrier Coverage

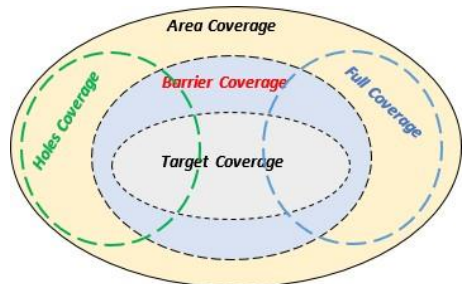

Figure 6. The New WSN Coverage Types

## 5. CONCLUSION

Several classifications focus on proposed solutions to ensure 1-barrier coverage and/or K-barrier in WSNs. In this paper, we divide barrier coverage into two axes; (a) deterministic-based barrier coverage model, (b) uncertain- based barrier coverage model. In each model, we classify barrier coverage according to deterministic and uncertain 2D and 3D environments. Depending on the application's need for continuous or intermittent coverage, we classify (1) linear and (2) nonlinear barrier coverage into strong and weak barriers in static or dynamic topologies. The gap between existing work and practical application needs to be considered in future research to address cur- rent and future needs, such as: Boolean barrier coverage, probabilistic barrier coverage, and fuzzy barrier coverage. Based on the requirements for monitoring the occurrence, importance and hazards of phenomena. In this paper, we mainly answer the following questions: "what new types of barrier coverage have emerged, how to deal with the problem of coverage of hazardous area boundaries and borders, what is the proposed solution for this, what are Current and future of this sacred task What were the challenges and, what deployment topologies were used…".

## ACKNOWLEDGEMENTS

The authors would like to thank everyone, just everyone!

**Authors**

Adda Boualem is Associate Professor, Djahida Taibi and Aroua Ammarare Master's students in Department of Computer Science at University of Tiaret, Algeria.

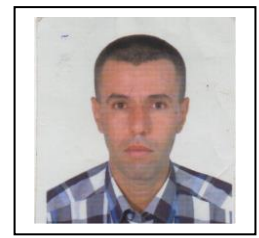